\pdfoutput=1

\documentclass[10pt]{article}

\usepackage{amsmath}
\usepackage{amssymb}

\usepackage{graphicx}

\usepackage{cite}

\usepackage{color} 

\usepackage[labelfont=bf,labelsep=period,justification=raggedright]{caption}

\makeatletter
\renewcommand{\@biblabel}[1]{\quad#1.}
\makeatother

\date{}

\usepackage[greek, ngerman, english]{babel}
\usepackage[latin1]{inputenc}
\usepackage{stmaryrd}

\begin{document}

\begin{flushleft}
{\Large
\textbf{Evolutionary distances in the twilight zone -- a rational
  kernel approach}
}
\\
Roland F. Schwarz$^{1,\ast}$, 
William Fletcher$^{2}$, 
Frank F\"orster$^{3}$,
Benjamin Merget$^{3}$,
Matthias Wolf$^{3}$,
J\"org Schultz$^{3}$,
Florian Markowetz$^{1,\ast}$
\\
\bf{1} CRUK Cambridge Research Institute, University of Cambridge,
Cambridge, UK
\\
\bf{2} Department of Genetics, Evolution and Environment and Centre
for Mathematics and Physics in the Life Sciences and Experimental
Biology, University College London, London, UK
\\
\bf{3} Department of Bioinformatics, Biocenter, University of
W\"urzburg, W\"urzburg, Germany
\\
$\ast$ E-mail: rfs32@cam.ac.uk, florian.markowetz@cancer.org.uk

\end{flushleft}

\section*{Abstract}
  Phylogenetic tree reconstruction is traditionally based on multiple
  sequence alignments (MSAs) and heavily depends on the validity of
  this information bottleneck. With increasing sequence divergence,
  the quality of MSAs decays quickly. Alignment-free methods, on the
  other hand, are based on abstract string comparisons and avoid
  potential alignment problems. However, in general they are not
  biologically motivated and ignore our knowledge about the evolution
  of sequences.  Thus, it is still a major open question how to define
  an evolutionary distance metric between divergent sequences that
  makes use of indel information and known substitution models without
  the need for a multiple alignment.
  Here we propose a new evolutionary distance metric to close this
  gap. It uses finite-state transducers to create a biologically
  motivated similarity score which models substitutions and indels,
  and does not depend on a multiple sequence alignment. The sequence
  similarity score is defined in analogy to pairwise alignments and
  additionally has the positive semi-definite property. We describe
  its derivation and show in simulation studies and real-world
  examples that it is more accurate in reconstructing phylogenies than
  competing methods. The result is a new and accurate way of
  determining evolutionary distances in and beyond the twilight zone
  of sequence alignments that is suitable for large datasets.

\section*{Author Summary}

\section*{Introduction}
State-of-the art phylogenetic reconstruction methods are currently
being challenged. For a long time, multiple sequence alignments
followed by maximum-likelihood (ML) tree reconstruction have been seen
as the computationally expensive gold standard for phylogenetic
analyses \cite{Whelan2001, Chor2005}. Distance approaches that base
their inference on summary statistics have traditionally been seen as
a fast but less precise alternative \cite{Kuhner1994}.
However, recent results point out that the gap between ML and distance
methods may be less pronounced than previously thought. For example,
the expected required sequence length for the reconstructed tree to
converge to the true tree phylogeny is not worse in distance-based
approaches than in ML \cite{Roch2010}.
Additionally the quality of the multiple sequence alignment heavily
affects reconstruction accuracy, a situation worsened by the
NP-hardness of the alignment problem and the heuristics used to cope
with it \cite{Goldman1998, Ogden2006, Wong2008, Wang2009,
  Fletcher2010}. The problem of alignment errors arises especially on
large-scale phylogenies with many taxa that span a broad divergence
range \cite{Talavera2007}, where many homologies lie in the
\emph{twilight-zone} of sequence alignments \cite{Doolittle1987}.

In the light of these findings, alignment-free distance-based
reconstruction methods deserve special attention, as they circumvent
potential pitfalls of the multiple alignment approach, especially with
respect to divergent sequences, and can be advantageous in speed
possibly without sacrificing reconstruction accuracy. Unfortunately
many purely alignment-free approaches \cite{Otu2003, Ulitsky2006}
lack unique biological motivation (for a comparison see also
\cite{Hoehl2007}). Joint estimation of
trees and alignments is computationally expensive
and relies heavily on heuristics and/or sampling approaches
\cite{Thorne1991, Thorne1992, Suchard2006, Rivas2008,Loeytynoja2008}.
The question of reconstructing phylogenies directly without multiple
alignment has only recently been tackled \cite{Daskalakis2010} with
promising results. We follow the basic principles of this approach but
here wish to present the phylogenetic reconstruction problem in a
different light.

Since there exists a one-to-one relationship between binary trees and
additive metrics \cite{Waterman1977} the phylogenetic problem of
finding the true tree is equivalent to finding the true additive
dissimilarity matrix.  Finding additive distances is hard, thus
distance-based approaches usually aim at finding a distance which is
as close as possible to the true additive one, so that the tree
reconstruction process which turns these non-additive distances into
additive trees finds the true tree as often as possible.  Metrics in
general, including additive distances, can be thought of as being
induced by a dot product $\left<\cdot, \cdot\right>$ in some Hilbert
space of possibly infinite dimension \cite{Schoelkopf2002}. Key to
phylogenetic reconstruction is constructing a Hilbert space and
associated dot-product such that distances between sequences are
indeed a measure of evolutionary divergence. Doing this explicitly is
impossible, if the space is of infinite dimension.  However, it can be
achieved implicitly by applying the so-called \emph{kernel-trick}
\cite{Schoelkopf2002}: A positive-definite (pd) kernel function
$k(\cdot, \cdot)$ in the input space (i.e. directly on the sequences
in our case) computes the scalar value of the dot-product in the
Hilbert space without explicitly constructing it.

The kernel trick has been applied successfully in a variety of
different fields, including natural language processing, face
recognition, speech recognition and computational biology.  Here we
extend its use to the problem of phylogenetic reconstruction. The
major challenge here is finding the right pd kernel. The pairwise
similarity measure between sequences must map sequences to an
evolutionary feature space ruled by the modification of sequences in
terms of insertions, deletions and substitutions. The natural distance
in this space should then come as close as possible to the true
evolutionary distance on the sequences.

In this article we derive such a kernel. Making use of classical
results from global pairwise alignment we show how a different
formulation of the alignment problem can map sequences to a feature
space of insertions, deletions and substitutions and gives rise to a
pd kernel. We study this similarity measure in its topological
reconstruction accuracy of phylogenetic trees from simulated and real
data. We show its superiority over conventional methods for
phylogenetic studies with a broad range of sequence divergence in and
beyond the twilight zone of remote homology. We further investigate
possible benefits of including suboptimal alignments into the score.

\section*{Materials and Methods}
Hidden Markov Models (HMMs) have been extensively used for
probabilistic modeling of sequence families, database searches and
other tasks.  Pair-HMMs work on two sequences simultaneously and are
capable of probabilistic modeling of pairwise alignments
\cite{Durbin1998}. The field of natural language processing uses
close relatives of HMMs, so called \emph{finite-state transducers
  (FST)}, for modeling the transformation of one sequence into
another or describing joint distributions on two sequences
\cite{Mohri1996}. Their advantage over pair-HMMs is the rigorous and
general definition which allows not only for probabilistic
interpretations, but for any set of values that follows specific rules
(more precisely all \emph{semirings}) to be used as weights. These
include, for example: probabilities, logarithmic numbers (where the
weights are summed along a path instead of multiplied); and boolean
values. In the following we use FSTs to create our kernel for
evolutionary sequence comparison.
We make use of two major observations: (i) The classical
problem of pairwise alignment can be posed as a shortest-path problem
on a log-weighted FST \cite{Mohri2003}; and (ii) FSTs that can be
decomposed into another FST and its inverse give rise to pd rational
kernels \cite{Cortes2004}.

\paragraph{Semirings}
The different classes of weights that can be used for FSTs are the
so-called semirings. They define two operations on a set, an abstract
sum and multiplication. In the case of the \emph{real} semiring, the
final score for two sequences is the (conventional) sum of all
possible paths generating those two sequences, where the weights in
each path are (conventionally) multiplied. Weights on the \emph{real}
semiring can be converted to the \emph{log} semiring by the link
function $\psi(x) = \exp(-x)$. In the \emph{log} semiring,
multiplication is turned into summation and the sum is replaced by the
logarithm of the sum. The \emph{tropical} semiring is a special
instance of the \emph{log} semiring in that the log-sum over all paths
is replaced by the \emph{minimum}, and corresponds to the
\emph{Viterbi} approximation in conventional HMMs. For a more formal
definition of semirings, see SI Text or \cite{Cortes2004,
  Mohri2009} and references therein.

\paragraph{Alignments Problems as FSTs}
Any edit-distance can be computed via a FST over the \emph{tropical}
semiring \cite{Cortes2004}. This includes the classical edit-distance
\cite{Levenshtein1966} as well as any generalized alignment
problem. The alignment score is then the minimum of all possible paths
of transforming one sequence into another. More formally, for a FST
$T$ over the \emph{tropical} semiring, the alignment score is defined
as $ \llbracket T \rrbracket(x,y) = \min_{\pi \in P(q,x,y,F)} \sum
w[\pi_i]$, where $P(q,x,y,F)$ is the set of all paths going from the
initial states $q$ to the final states $F$ thereby transforming $x$ to
$y$. The standard global pairwise alignment problem for example is a
three state FST with a map, an insert and a delete state. The
self-transitions in the match state are weighted with the scores of
the used substitution matrix, the transitions to the gap states and
the self-transitions in the gap state are weighted with the gap open /
gap extend costs respectively.

\paragraph{PD Rational Kernels and distances}
A FST $T$ over the \emph{real} semiring associates a real-valued
number with every pair of sequences $(x,y)$. This score is then the
\emph{sum} over all possible paths transforming $x$ to $y$,
\emph{multiplying} instead of summing the weights along the path. This
mapping from two-tuples of the space of sequences to the reals is
called a rational kernel. If the transducer $T$ can further be
decomposed into a transducer $S$ and its inverse $S^{-1}$ ($T = S
\circ S^{-1}$), the kernel is known to be pd \cite{Cortes2004} (for
details on FST composition and inversion see SI Text or
\cite{Mohri2009}). In this setting, the transducer $S$ performs the
feature space mapping. It encodes the prior knowledge about the
features important for our problem domain. From a pd kernel we can
directly compute distances in the feature space via $d(x,y) = \|x -
y\|^2_k = k(x-y, x-y) = k(x,x) - 2k(x,y) + k(y,y)$. 

\paragraph{Pairwise alignments as pd rational kernels}
It is our goal to modify the pairwise alignment problem in a way that
we can prove the resulting alignment score to be pd. To achieve this
we replace the min operation by the log-sum, thereby changing
semirings from the tropical to the log. The resulting score includes
all possible (suboptimal) alignments. By making use of the link
function $\psi(x) = \exp(-x)$ we can convert that logarithmic score
into a real value. The result is the score of a rational kernel
\cite{Cortes2004}.

\begin{figure*}[!ht]
  \centerline{\includegraphics[width=\textwidth]{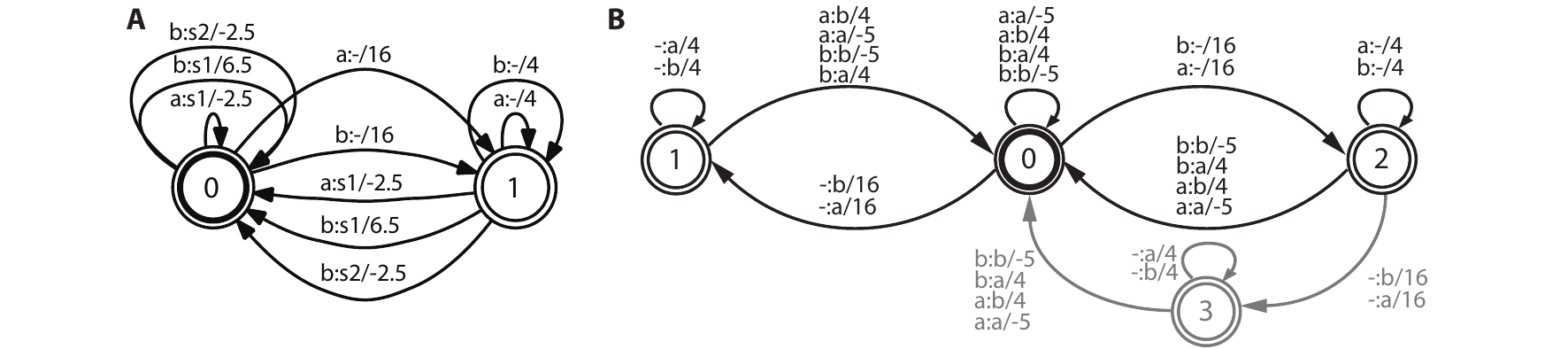}}
  \caption{\label{fig1} \textbf{A: Feature space mapping for
    biological sequences} using a FST over the log semiring: Every
    transition has an attached input and output symbol separated by a
    colon, and an associated weight. Symbols can either be kept,
    substituted or deleted. Composition of such a transducer with its
    on inverse yields a pd rational kernel. The alphabet has been
    reduced to two symbols for illustration purposes, $-$ depicts a
    gap or epsilon transition. \textbf{B:} Result of the composition
    of the transducer encoding the feature space mapping with its
    inverse: The starting state (state 0) corresponds to the match
    state, the additional two colored states (states 1, 2) encode
    insertion and deletion states. The transitions to the gap states
    are scored with gap open costs and the self transitions in the gap
    state with gap extend costs. The additional fourth state (3) is a
    result of the epsilon filtering process during composition.}
\end{figure*}

To see that this kernel is indeed pd we need to decompose it in to a
feature space mapping FST and its inverse $T = S \circ S^{-1}$. On the
real semiring and ignoring epsilon transitions (gaps), it is easy to
see that by the definition of composition this equals a Cholesky
Decomposition of the transition weight matrix, which requires the
pointwise exponential of the substitution matrix used to be pd.  If we
wish to include gaps we need to construct the feature space
explicitly:

We can think of a feature space mapping where each position in a
biological sequence can either be retained, substituted or deleted
using some intermediate alphabet $s_1 \dots s_n$. For an example of
such a FST with weights derived from a standard nucleotide
substitution matrix and gap scores of $16/4$ see Figure
\ref{fig1}a. Composition of this FST with its own inverse, obtained by
reversing input and output symbols, leads to the FST in Figure
\ref{fig1}b.

It can easily be seen how the composed FST again resembles the
topology of a global alignment FSA \cite{Durbin1998}, with a match
state and two states corresponding to insertions or deletions. The
additional fourth state contained in the transducer is a result of the
epsilon filter used. Different epsilon filters lead to different
topologies \cite{Mohri1996, Mohri2009} where the three-state backbone
of match, insertion and deletion states are always retained. This
additional path theoretically allows for the opening of a new gap
within a gap, something which is automatically excluded if looking for
the shortest path or best-scoring alignment between two sequences.

In summary, reformulation of the classical global pairwise alignment
paradigm allows for the interpretation of the alignment score as a
shortest-path approximation of the kernel score of a pd rational
kernel working on biological sequences. 

\paragraph{The impact of suboptimal alignments on the kernel score}
If the absolute difference between the summands of a logarithmic sum
is large the sum is heavily dominated by its smaller
summand. Therefore, in cases where the optimal alignment score is
distinctively smaller than any suboptimal alignment the kernel score
including all suboptimal alignments will be close to the shortest-path
approximation. In cases where even the best alignment score is not
significantly smaller than its closest suboptimal siblings the full
score will differ. In order to be able to study the effect of the
inclusion of suboptimal alignments in terms of reconstruction accuracy
we project the exponential of the matrix of pairwise alignment scores
to the next positive semi-definite \cite{Higham1988}. This
shortest-path approximation is not neccesarily pd anymore. How big the
difference is depends on the optimality of the best score as discussed
above.

\section*{Results}
We performed repeated simulation experiments to validate our distance
measures using nucleotide and amino acid sequences over two different
tree topologies and on each with increasing sequence divergence.
\paragraph{Sequence simulation}
Amino acid and nucleotide sequences were generated according to two
tree topologies with 18 and 52 taxa in realistic scenarios using
INDELible \cite{Fletcher2009} (see also SI text). Trees were
reconstructed and topologically compared to the true tree using the
quartet distance \cite{Mailund2004}. The studied methods were (i)
traditional multiple alignment using \emph{Muscle} \cite{Edgar2004}
followed by Jukes-Cantor distance estimation using \emph{Phylip}
\cite{Felsenstein2009}, (ii) statistical consistency alignment using
\emph{ProbCons} based on pair-HMMs \cite{Do2005} followed by
\emph{RAxML} maximum-likelihood tree reconstruction
\cite{Stamatakis2006}, (iii) an alignment-free method of distance
estimation based on the Lempel-Ziv complexity \cite{Otu2003}, (iv) a
pattern-based maximum-likelihood approach for alignment-free distance
estimation \cite{Hoehl2006} and (v) the classical Levenshtein
distance \cite{Levenshtein1966}. Comparison according to (iv) had to
be performed on a much smaller sample size due to the high
computational demand of the method \cite{Hoehl2007}. In a preliminary
study we found (iv) to perform only slightly better than (iii) for
closely related sequences. We thus kept method (iii) as a
representative for alignment-free methods.

\begin{figure*}[!ht]
  \centerline{\includegraphics[width=\textwidth]{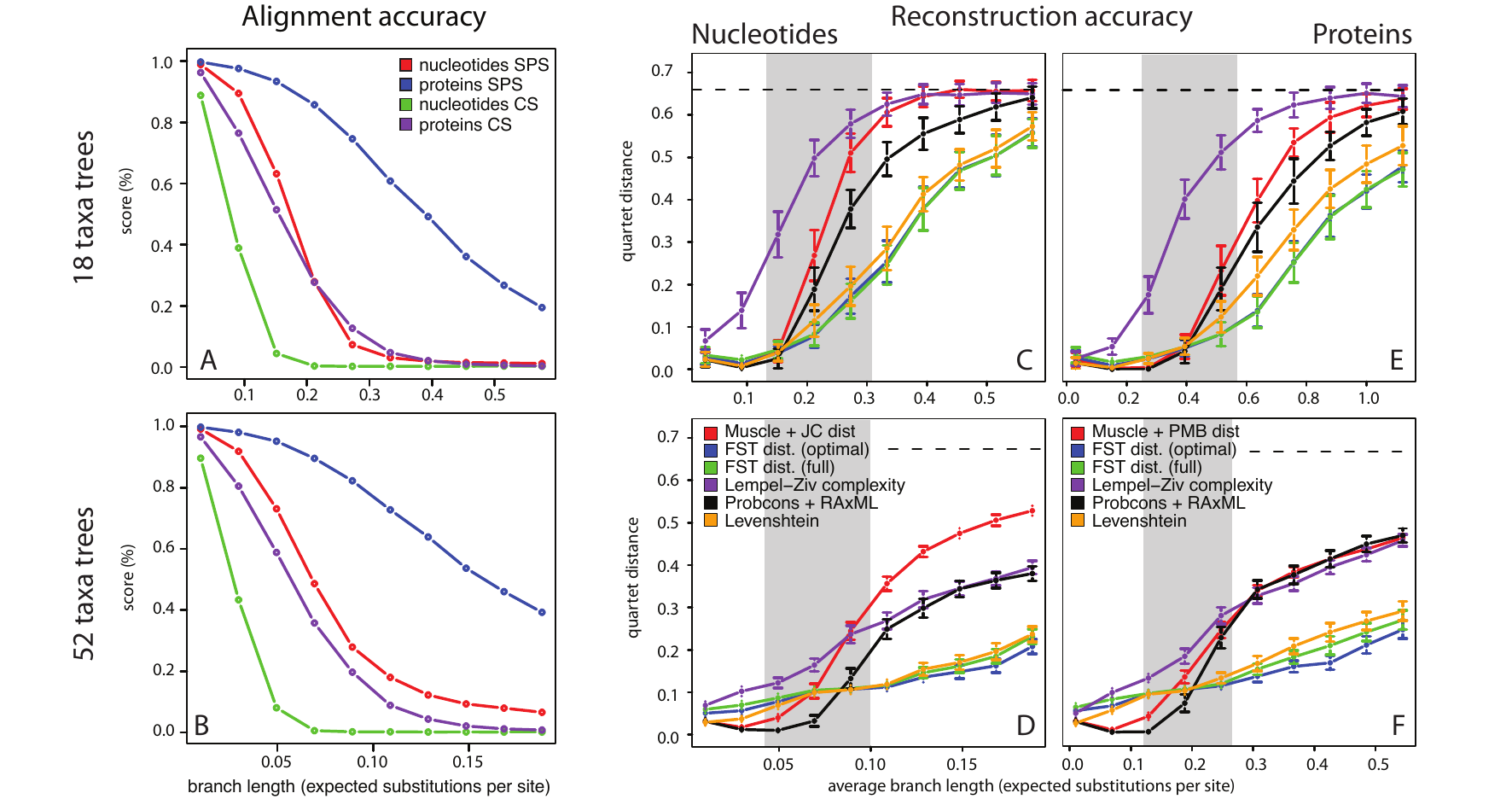}}
  \caption{\label{fig2} \textbf{A,B: Alignment accuracies measured in
    column and sum-of-pairs scores}. With increasing branch lengths
    multiple alignments accumulate errors, which leads to the poor
    reconstruction accuracies observed. \textbf{C-F: Simulation
    results for nucleotide sequences (left) and protein sequences
    (right)}: All experiments were repeated 100 times, standard error
    estimates are shown. The traditional approach of multiple
    alignment followed by distance estimation performs well for
    closely related sequences (red and black lines). The error curve
    of the two FST approaches has a significantly lower slope and
    performs well even for divergent sequences (green and blue), so
    does the classical edit-distance (yellow) which is still behind
    the FST distances. Statistical consistency aligners (black)
    perform better than traditional aligners (red) but suffer from the
    same rapid decay in reconstruction accuracy. The Lempel-Ziv
    complexity-based distance only achieves good results for the 52
    taxa tree (purple). The dotted black line at the top gives the
    maximum expected quartet distance from a random tree.}
\end{figure*}

\paragraph{Sequence divergence leads to poor alignment quality}
To assess the impact of sequence divergence on multiple alignment
accuracy we first compared the alignments from Muscle with the true
INDELible alignments. We calculated two scores to quantify this
accuracy (Figure \ref{fig2} A,B): The \emph{column score} (CS) is the
proportion of columns from the true alignment that are present and
correct in the test alignment. The \emph{sum-of-pairs score} (SPS) is
the proportion of aligned pairs of nucleotides/amino-acids from the
true alignment that are also aligned together in the test
alignment. The first is a very stringent measure as all
nucleotides/amino-acids in a column must be correctly placed for that
column to be deemed correct. The latter is a more lenient measure as
it rewards correct alignment between some sequences even if other
sequences in that column are mis-aligned \cite{Fletcher2010}.

The results show that the number of correctly aligned positions
exponentially decreases with increasing sequence divergence for
nucleotide sequences. Amino acid sequences showed a more linear trend,
possibly due to the higher information content introduced by the
larger alphabet size of amino acids as compared to nucleotides, but
suffer from the same effect.

\paragraph{FST distance for divergent sequences}
Quartet distances between the estimated and true trees for nucleotide
as well as amino acid sequences over all tree topologies
(Figure~\ref{fig2}, C-F) show that the traditional approach of a
multiple alignment followed by distance estimation is highly accurate
for closely related species. When entering the twilight zone of
sequence alignments reconstruction accuracy drops exponentially. Above
average branch lengths of $> 0.1$ substitutions per site for the 52
taxa nucleotide tree and $> 0.4$ substitutions per site for the 18
taxa protein tree the multiple alignment becomes erroneous (red and
black lines) and tree reconstruction accuracy gets weak as the number
of quartets in common with the true tree approaches that of a random
tree (dotted line). This effect is about $3$ times stronger for
nucleotide then for protein trees and about $2$ times stronger in the
52-taxa tree as compared to the 18-taxa tree. To exclude
aligner-specific artifacts, we included ProbCons (black line, Figure
\ref{fig2}) into the analysis. We additionally included ClustalW which
was found to perform slightly worse than Muscle. We also computed
maximum-likelihood trees on the alignments to assess the difference in
reconstruction accuracy between simple distance-based and
character-based approaches. The RAxML trees outperformed the distance
based trees by a margin which was more profound for amino acid
sequences than for nucleotide sequences, but still suffered drastic
loss in reconstruction accuracy with increasing sequence divergence,
due to the accumulation of alignment errors. The \emph{ProbCons}
results showed that aligners specifically designed to address these
issues indeed perform slightly better to moderately better across all
experiments but still suffer from a rapid loss in reconstruction
accuracy with increasing sequence divergence. The alignment-free
methods generally performed worse than other methods tested on the 18
taxa tree but were close to the best alignment-based methods for the
52 taxa trees.

Distance estimation using our proposed finite-state transducers (blue
and green lines) came close to the performance of classical multiple
alignment for closely related species. It showed only a gradual
decrease in reconstruction accuracy with increasing evolutionary
divergence, being significantly more accurate than any other method
tested. This evidently shows that the classical approach of multiple
alignment followed by tree reconstruction is superior only if the
alignment is correct. From a certain distance on, multiple sequence
alignments cannot be reconstructed accurately any more, leading to
poor reconstruction power in the downstream phylogenetic
analysis. This seems to hold for classical progressive multiple
alignment as well as statistical consistency alignment.

\paragraph*{The influence of suboptimal alignments on the kernel score}
When comparing both proposed kernel scores, one incorporating all
suboptimal alignments into the score, the other only using the optimal
alignment, we noticed differences between the two tree topologies: In
the 18-taxa case both variants perform equally well. In the 52-taxa
case the suboptimal alignments added more noise to the score than
signal and the kernel using only the optimal score came out
ahead. Even though the average branch length in the 52 taxa tree is
moderate the variance of pairwise distances between sequences is
higher when the tree contains more branches. When comparing sequences
beyond the twilight zone the scores of the optimal and subsequent
suboptimal alignments are similar, influencing the kernel score (see
\emph{Methods} section). The good performance of the score based on
the optimal alignment is especially attractive as this shortest-path
approximation can be computed with standard global alignment
implementations, such as the Stretcher program from the EMBOSS package
followed by projection to the next pd matrix. We were already able to
test this procedure in a real-world application, comparing 500 human
kinases with 2600 kinases from \emph{Paramecium tetraurelia}. We
showed that the kinome of \emph{P. tetraurelia} is more than 5 times
the size of the human kinome. In addition to whole genome
duplications, further duplications lead to the expansion of specific
subfamilies. More than 20 ciliate specific domain architectures were
discovered \cite{Bemm2009}.

The most traditional way of pairwise comparison between sequences is
the edit-distance or Levenshtein distance
\cite{Levenshtein1966}. Results show, that like our own pairwise
sequence comparisons the Levenshtein distance is not prone to the
multiple alignment pitfall and therefore performs well for divergent
sequences. It provides a relatively accurate estimator for distances
between nucleotide sequences but performs worse for protein
sequences. This is not surprising as the EDNAFULL matrix traditionally
used for nucleotide alignments scores matches with $5$ and all
substitutions equally with $-4$. The information content of this
matrix is not higher than that of the edit-distance matrix which
scores matches with $0$ and any edit operation with $-1$. The picture
changes in the case of protein substitution matrices that carry dense
information about the exchangeability of amino acids.

\paragraph*{FST distance places \textit{Sphaeroplea} clade correctly
  without information about secondary structures}
We applied our method to a set of 52 internal transcribed spacer II
(ITS2) sequences of the Chlorophyceae \cite{Keller2008}. The group
consists of 6 major clades of high within-group sequence similarity
($90\%$ median pairwise sequence identity) but significant divergence
between groups (down to $67\%$ total average sequence identity).  Over
the last few decades there has been ongoing discussion about placement
of the \textit{Sphaeroplea} clade within this set of taxa
\cite{Buchheim2001, Wolf2002, Mueller2004}. Even though most authors
agree on the existence of a monophyletic \emph{DO}-group comprising
the \textit{Sphaeroplea}, \textit{Hydrodictyon} and
\textit{Scenedesmus} clades, the position of the \textit{Sphaeroplea}
clade within this group was only recently verified by taking
structural properties of the ITS2 into account \cite{Keller2008}.

\begin{figure*}[!ht]
  \centerline{\includegraphics[width=\textwidth]{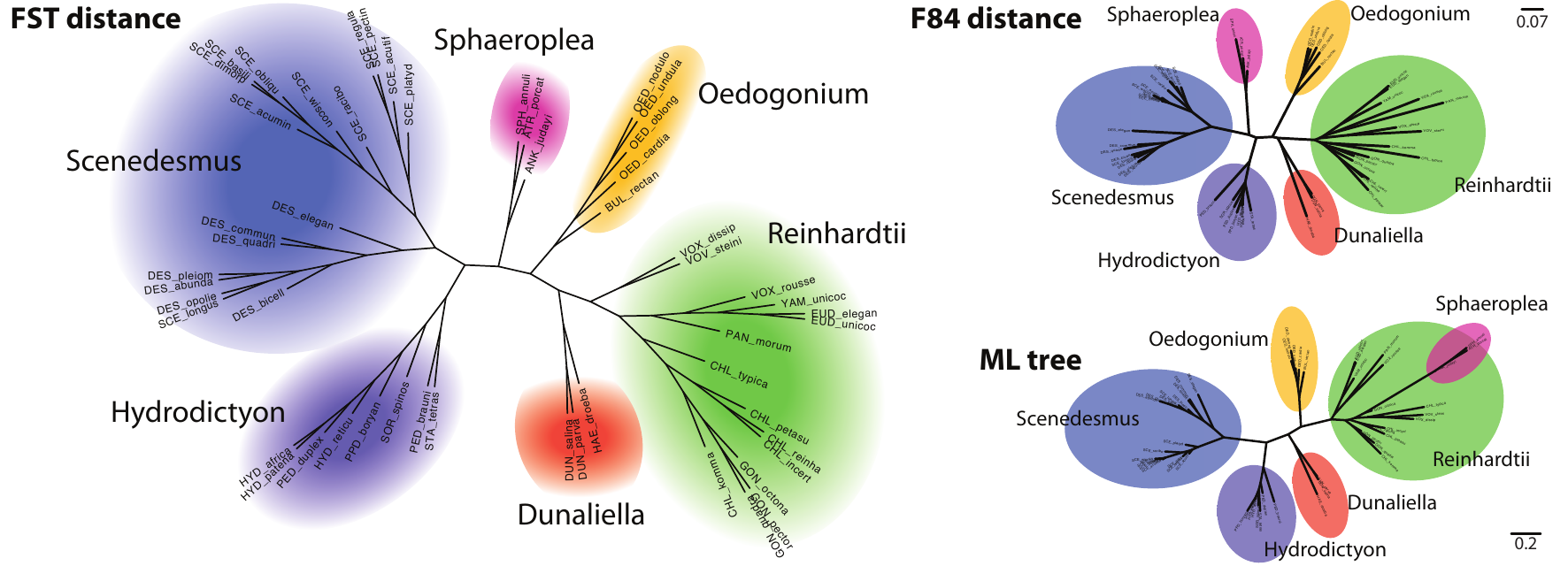}}
  \caption{\label{fig3} \textbf{Reconstructed phylogenetic trees of the
    Chlorophyceae group} for three different methods: FST distance
    (left) using the full kernel score, F84 distance estimation on a
    Muscle alignment (top right) and maximum-likelihood tree on the
    same Muscle alignment (bottom right). Only the FST tree reveals
    the same grouping of the major clades as discussed in
    \cite{Keller2008}, which we use as a `gold standard'.  The distance tree erroneously places the
    \textit{Sphaeroplea} clade between the \textit{Hydrodictyon} and
    \textit{Scenedesmus} clades, while the ML tree places them within the
    \textit{reinhardtii}-subgroup.}
\end{figure*}

We applied our FST distance method to the set of ITS2 sequences and
compared it to both classical distance estimation and maximum
likelihood tree reconstruction on a multiple sequence alignment
(Figure \ref{fig3}). Comparing the MSA with the manually curated
sequence structure alignment taken from \cite{Keller2008} shows that
the Muscle alignment contains many misaligned columns (CS $0.096$, SPS
$0.59$).  The reconstructed trees differ from the true tree,
especially with respect to the placement of the \textit{Sphaeroplea}
clade. The distance tree places the \textit{Sphaeroplea} clade between
the \textit{Hydrodictyon} and \textit{Scenedesmus} clades (Figure
\ref{fig3} right top), whereas the ML tree again places the
\textit{Sphaeroplea} clade within the \textit{reinhardtii}-subgroup
(Figure \ref{fig3} right bottom). Our FST distance, which circumvents
the multiple alignment step, correctly places the \textit{Sphaeroplea}
clade next to the \textit{Hydrodicton/Scenedesmus} sister clade.  In
other methods this position can only be inferred by using additional
secondary structure information to reduce alignment errors.  Our
method was additionally able of correctly grouping a monophyletic
\textit{Gonium} clade.

\section*{Discussion}
In this paper we have shown that a kernel-based distance measure
circumvents problems of MSA quality and performs very well in and
beyond the twilight zone of remote homology. We intentionally used
known substitution matrices and gap scores as parameters to illustrate
the link to classical global alignments. Custom parameters estimated
by e.g. expectation maximization over alignments of a given divergence
range will supposedly perform even better.

Using FSTs to derive the distance has several advantages. For example,
the inputs to the distance calculation are currently two individual
sequences, formulated as finite-state acceptors that emit exactly the
sequence under study. This can seamlessly be extended to acceptors
emitting distributions over sequences, i.e. Hidden Markov Models like
profile-HMMs \cite{Eddy1998} to compute distances between sequence
families with possible applications in e.g. Profile-Neighbor-Joining
\cite{Friedrich2005}. The construction of a pd kernel using
composition of two individual FSTs is a necessary step, as generalized
edit-distances like the classical pairwise alignment score, are not
negative definite and therefore can not easily be turned into a pd
kernel by exponentiation alone \cite{Cortes2004}.

The methods we compare ourselves against are amongst the most
frequently applied, such as \emph{Muscle} followed by a distance-based
tree reconstruction, but also include state-of-the art statistical
consistency aligners like ProbCons followed by ML tree
reconstruction. As both JC distance and ML reconstruction methods
suffer from the same decrease in accuracy we show that it is not
simply the modeling of insertions and deletions that is improved in
our distance measure as compared to JC. We additionally clustered
sequences by length to see if the sheer number of insertions and
deletions in divergent sequences were mainly responsible for this
effect. This clustering performed very poorly as expected.

Sophisticated methods of statistical alignment \cite{Bishop1986} are
capable of computing joint probabilities for sequence comparison, but
the derivation of distances remains arbitrary to a
degree. Furthermore, such statistical methods can be found lacking
because of simplifications such as assuming that indel events involve
only one residue (TKF91 model, \cite{Thorne1991}) or that sequences
are made from non-overlapping indivisible fragments (TKF92 model,
\cite{Thorne1992}). Maximum-likelihood estimates for the time elapsed
between two species given the sequences additionally involve
reversibility assumptions and solving non-convex optimization problems.

Other algorithms, such as \cite{Hein2003}, are only
practical in analyses involving a small number of sequences. They
necessarily need to be coupled to numerical optimization methods
to find maximum likelihood estimates of parameters such as insertion
and deletion rates, substitution parameters, and branch lengths. In contrast, 
our approach is capable of directly using substitution matrices that
are known to perform well for certain evolutionary distances.

In summary, the present fast and MSA-free methodology allows us to
compute pairwise distances between sequences that mirrors the global
pairwise alignment process.  Our methodology interprets alignment
scores as values of a kernel that implicitly maps sequences to a
feature space with a biologically motivated topology: it is built of
modifications of that sequence using insertions, deletions and
substitutions.  Our methodology can directly be applied to compute
distances between distributions of sequences.  The resulting pd kernel
matrix can be used in any method that can be expressed in terms of dot
products alone (e.g. classification via support vector machines).
The distances are meaningful in evolutionary terms and outperform
other phylogenetic inference methods on divergent sequences in and
beyond the so-called twilight zone of remote homologies.  Thus, our
methods complement traditional approaches for more closely related
sequences.
Future work will focus on assessing the robustness of the kernel score
(bootstrapping) and the question of mapping sequences directly to an
additive space, i.e. from which additive distances can be immediately
derived, to remove the final approximation step when going from the
matrix of pairwise distances to the tree.

\section*{Acknowledgments}
The authors thank Cyril Allauzen for helpful comments and answers
regarding the finite-state transducer framework per mail as well as
on the OpenFST forum.

\bibliography{transducer}

\section*{Figure Legends}

\newpage

\newpage



\end{document}